\documentclass[preprint,5p,number,sort&compress]{elsarticle}
\usepackage{dcolumn}
\usepackage{bm}
\usepackage{textcomp}
\usepackage{upgreek}
\usepackage{graphicx}
\usepackage{hyperref}
\usepackage[load-configurations=abbreviations,detect-all=true]{siunitx}
\usepackage[version=3]{mhchem}
\usepackage[mathlines]{lineno}
\usepackage[usenames,dvipsnames,svgnames,table]{xcolor} 
\usepackage[normalem]{ulem} 
\usepackage{transparent}

\usepackage{xspace}

\newcommand{\NP}{\emph{Q-Pix}\xspace}
\newcommand{\TL}{\textbf{\emph{tile}}\xspace}
\newcommand{\RS}{\emph{reset}\xspace}

\newcommand{\TMIIm}{\mbox{\emph{Topmetal-II\raise0.5ex\hbox{-}}}\xspace}

\DeclareSIUnit\keV{keV}
\DeclareSIQualifier\ee{ee}
\DeclareSIQualifier\nr{nr}
\DeclareSIUnit\pe{p.e.}

\newcommand{\sym}[1]{\texttt{#1}}
\newcommand{\etal}{et\penalty50\ al.}

\journal{\phantom{Nuclear Instruments and Methods in Physics Research A}}

\begin{document}

\begin{frontmatter}

\title{\emph{Q-Pix}: Pixel-scale Signal Capture for Kiloton Liquid Argon TPC Detectors: Time-to-Charge Waveform Capture, Local Clocks, Dynamic Networks}

\author[uta]{David Nygren\corref{cor0}}\ead{nygren@uta.edu}
\address[uta]{University of Texas at Arlington}
\cortext[cor0]{Corresponding author}
\author[lbl]{Yuan Mei}
\address[lbl]{Lawrence Berkeley National Laboratory}

\begin{abstract}
We describe a novel ionization signal capture and waveform digitization scheme for kiloton-scale liquid argon Time Projection Chamber (TPC) detectors.  The scheme is based on a pixel-scale self-triggering `charge integrate/reset' block, local clocks running at unconstrained frequencies and dynamically established data networks.  The scheme facilitates detailed capture of waveforms of arbitrary complexity from a sequence of varying time intervals, each of which corresponds to a fixed charge integral.  An absolute charge auto-calibration process based on intrinsic \ce{^{39}Ar} decay current is a major benefit.  A flat electronic architecture with self-guided network generation provides exceptionally high resilience against single-point failure.  Integrated photon detection, although highly speculative, may also be possible.  The goal is optimized discovery potential.  Much might be at stake.

\end{abstract}

\end{frontmatter}

\section{Introduction}\label{sec:intro}
\subsection{Context}

The DUNE project is the flagship enterprise of US HEP\cite{Acciarri:2016crz}.  Definitive results are expected for neutrino mass ordering, with high sensitivity to CP-violation in the lepton sector, supernova dynamics, other neutrino topics and the search for baryon decay.  The DUNE neutrino physics program depends on many things, but accurate classification of events is the one big thing: is the apparent single electron in this event really caused by neutrino flavor change, or is that track a misidentified $\pi^0$ or $\gamma$?  For CP-violation studies, discrimination among various neutrino flavor interactions and nuclear/hadron/electron recoils is of paramount importance.  In the end, after several years of running, the number of true events is likely to be measured in tens to a few hundred.

Many processes of fundamental interest lead to events that display complex signal waveforms with highly variable features in space, amplitude and time.  As the intrinsic quality of information inherent in a kiloton liquid argon TPC (LArTPC) is very high, the central technical goal must be to capture these events with high efficiency, and in exquisite detail.  High quality data may ultimately contain surprises whose significance may exceed what is now foreseen.

The massive scale of the DUNE far detector (FD) modules invites compromises in signal capture and data quality.  But are such compromises really necessary?  Will any unexpected phenomenon, however subtle but above intrinsic sensitivity threshold, be recognized as new physics, be dismissed as artifact, or just not seen? We seek here to optimize discovery potential.

In the era of DUNE operation as a full system, however, much more will have become known about primary goals from other experiments.  The mass ordering of light neutrinos is likely to have been well clarified, and no follow-on experiment based on a particular value of CP-violation is apparent.  The enormous investment, with sustained effort for two decades by thousands of people, is likely to exceed US\SI{2e9}[\$]{}.  From this mixed perspective, how should the DUNE program be viewed?

We believe an obligation exists within the HEP community to ensure that data extracted from the DUNE FD is captured with the highest quality to the extent possible within constraints of time, resources, and our capacity to innovate, so that any subtle new phenomena near the limit of detection may be revealed.  A search for an optimal, intrinsically 3D signal capture approach with low signal threshold seems accordingly very well motivated.  Dwyer \etal\cite{Dwyer:2018phu} have recently demonstrated a true 3D pixelized information capture technique, proposed for the DUNE near detector.  That work provides evidence that impressive performance is possible, and is thus a seminal advance.

Our pixelization concept is aimed toward the DUNE FD.  While this goal may seem impractically ambitious, that perception may not be true.  When a true signal event occurs, high-quality capture of all spatial and topological information, energy loss, and event time is desired over a huge addressable volume-time space.  However, almost all the time, nothing interesting happens.  The circuitry must hence do as little as possible yet respond properly to the arrival of signal.  This very low rate of intrinsically interesting signals can be exploited to technical advantage.  At the same time, the technical approach must display operational availability well above \SI{99}{\percent} for a decade and be extremely resilient against single-point failure.

Given the challenges here, any complete solution---should any exist---will likely appear unorthodox.  In stark contrast to tradition, our concept, \NP, captures waveforms by measuring time per unit charge.  \NP is thus a radical departure for information capture, but one that arguably may provide an enabling technical advantage for discovery.

\subsection{Information quality in time and space}

Reconstruction accuracy is limited in large measure by diffusion of the drifting electrons, up to $\sim\SI{4}{mm}$ rms transverse and $\sim\SI{14}{mm}$ rms longitudinal at the maximum drift distance, which is taken here to be $\sim\SI{7}{m}$.  Track-pair separation and practical minimum signal size suggest a pixelization scale of $4\times\SI{4}{mm^2}$.  This choice is likely near a soft optimum.  Electromagnetic showers, with scattered point-like energy depositions and numerous gaps of a few \si{mm} along apparent tracks, etc., should be reliably detected.

With $4\times\SI{4}{mm^2}$ pixels, there are \num{62500} pixels per \si{m^2} and more than \num{e8} addressable channels per \SI{10}{kTon} module (but not so different from an LHC pixelized vertex detector\cite{alice:its:tdr2014}).  The 3D voxel size is on the order of $4\times4\times\SI{4}{mm^3}$, corresponding to a granularity exceeding \num{e11}.

If a prompt scintillation signal \sym{S1} is available and can be associated with a particular track, time resolution better than \SI{0.05}{\micro s} is possible.  Assuming a typical drift velocity of \SI{1.6}{mm/\micro s}, placement along the drift coordinate is accordingly sub-\si{mm}. However, if \sym{S1} signals are not reliably detected and associated, placement along the drift coordinate is problematic.  The alternative is then to make highly accurate measurements of longitudinal and transverse diffusion and of signal attenuation to provide an approximate measurement of drift distance.

A LAr TPC can provide highly accurate, detailed track profiles, and \NP is intended to provide the desired measurement quality.  A muon with \SI{1}{GeV} energy will have track profiles measured about \num{1500} times, leading to statistically very precise measurements of diffusion during drift.  This capability could be of critical importance.

For SN events, data rates become very high for one second or so, with a range of deposited energy up to $\sim\SI{25}{MeV}$.  The longest tracks are hence at most a few \si{cm}; perhaps the high level of track detail captured by \NP can yield a sense of electron directionality.  In our scenario, local buffer depth is deep enough to capture all data from nearby (but not too close!) SN.

\subsection{Resilience and reliability}

The detection elements within LAr must be considered inaccessible during the entire lifetime of the experiment.  We take as a fundamental axiom that, for any design to be considered, it must display very robust resilience against Single Point Failure (SPF).  This coupled issue of reliability and inaccessibility was successfully confronted in IceCube, which deployed \num{5000} complex Digital Optical Modules (DOM) permanently in deep Antarctic ice.  The DOM ensemble has demonstrated excellent overall reliability in deployment: $\sim\SI{98}{\percent}$ viability, as well as essentially \SI{100}{\percent} operational availability.  Of the $\sim\SI{2}{\percent}$ losses, half ($\sim\SI{1}{\percent}$) is due to cable connector failure.  We take this successful experience as a guidepost in the development of our concept.  We assert that it is reasonable to deploy all signal capture electronics permanently within the FD LAr, for the entire lifetime of DUNE.

\section{\NP: time-to-charge ionization signal capture}
\subsection{The Charge-Integrate/Reset Circuit}

Waveforms of arbitrary complexity and wide dynamic range must be captured and time-stamped.  For pixels, the classic approach of continuous waveform sampling and digitization is clearly inappropriate.  In design, we must follow, in spirit and in practice, an electronic equivalent of the principle of `Least Action'.  How, then, to be normally `OFF' but then instantly `ON'?

\begin{figure}[!htb]
  \centering
  \includegraphics[width=0.8\linewidth]{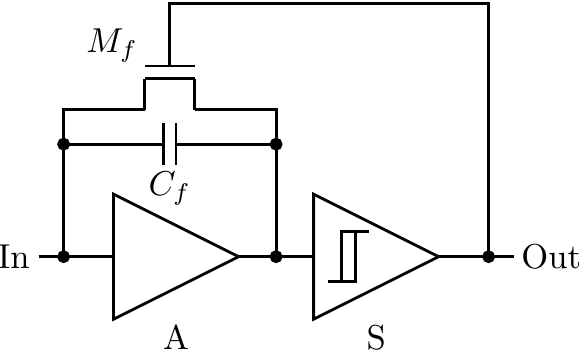}
    \caption{A symbolic representation of the charge integrator/reset is illustrated.  For clarity, polarities, power, ground, biases, thresholds, etc., are not shown.  A charge-sensitive amplifier, on the left, drives a regenerative comparator, on the right.  When the amplifier has integrated charge sufficient to present a voltage above comparator threshold, the regenerative comparator switches rapidly and completely due to positive feedback.  This transition \RS{}s the amplifier and the comparator switches back.  The sequence produces a short standardized \RS signal.}
  \label{fig:CSArstSchmitt}
\end{figure}

An unorthodox but surprisingly natural overall solution begins with the simple \textbf{Charge-Integrate/Reset} (CIR) circuit block, as shown in Fig.~\ref{fig:CSArstSchmitt}.  This approach is combined with a less familiar but now well-proven time-stamping scenario based on free-running local clocks, as proven in IceCube\cite{Aartsen:2016nxy}.

In \NP, a charge sensitive amplifier (CSA) continuously integrates incoming signal on a feedback capacitor until a threshold on the regenerative comparator (a Schmitt trigger) is met.  When that threshold is met, the comparator initiates a `\RS' transition that rapidly drains the feedback capacitor and returns the circuitry to stable (arbitrary) baseline.  The cycle then begins again, \emph{ad infinitum}.

Older, truly ancient implementations of this type of circuit were used to integrate macroscopic currents over long times by using the \RS pulse to increment a mechanical or electrical counter.  The `charge counter' sum is a measure of total integrated charge since counting began.

Our scenario is different from the original `charge counter' discussed above.  Here, the \RS transition pulse is used to capture and store the current value of a local clock within one ASIC that serves a group of 16 (or perhaps up to 32) pixels.  In \NP some current is lost during the short `\RS' transition but this is likely a small manageable correction.  The charge loss during \RS can also be minimized with additional elements in circuit design (see discussion below).

We acknowledge that the  engineering design challenges, to operate multiple sensitive analog input circuits with proximate active digital circuitry, are formidable.  Nevertheless, this situation has been successfully confronted in (Dwyer\cite{Dwyer:2018phu}) and other small-signal applications.

\subsection{Minimum-Ionization Signals}

From (Dwyer\cite{Dwyer:2018phu}) the most probably energy loss for a Min-I particle in LAr is about \SI{1.66}{MeV/cm}.  Taking the average energy to produce a free electron-ion pair in LAr to be \SI{33}{eV}, about \num{5e4} electrons comprise the average signal per \si{cm}.  For a $4\times\SI{4}{mm^2}$ pixel, a typical charge of an arriving track segment parallel to the readout plane is at least \num{20000} electrons, or $\sim\SI{3.3}{fC}$.  Of course charge varies considerably with angle of incidence and/or closest distance of approach to pixel center, etc.

\subsection{Dynamic range}

We suspect the actual track dynamic range requirement does not exceed about 30.  $dE/dx$ ranges from the predominant minimum-ionizing muons and electrons, to more ionizing tracks such as protons or $K^+$, etc. (perhaps from baryon decay in argon nuclei), and finally up to maximally ionizing nuclear fragments.  Measured energy and $dE/dx$ for the most strongly ionizing `tracks' will be strongly affected by nuclear effects and columnar plus volume electron-ion recombination.  While it seems likely to us that clean spatial reconstruction is far more important than precise measurements of $dE/dx$ for highly ionizing nuclear fragments, this question awaits careful simulation studies.  The given dynamic range and precision requirements\cite{Acciarri:2016crz} are intertwined and surely await further clarification.

\section{Basic information datum: reset time differences}

What we can exploit here is something new: the essential quantum of information for each pixel is the time difference between one clock capture and the next sequential capture: the \textbf{Reset Time Difference (\sym{RTD})}.  The \sym{RTD} measures the time to integrate a pre-defined integrated charge: $\Delta Q$.

\begin{figure}[!htb]
  \centering
  \includegraphics[width=0.8\linewidth]{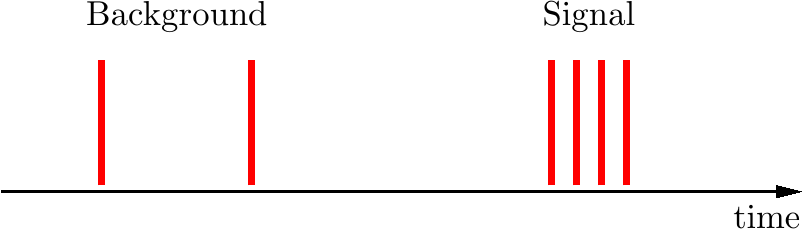}
    \caption{Illustration of inverse relationship between \RS intervals and current.  The small background current takes many \textbf{seconds} to initiate a \RS whereas signal currents lead to a sequence of short \RS intervals on a sub-\textbf{\si{\micro s}} scale.  Of course an illustration cannot display this important feature accurately on a linear scale.}
  \label{fig:RTD}
\end{figure}

\sym{RTD}s are not generated locally within the LAr, but by surface computing systems that perform subtractions in sequence from the string of \RS times associated with a particular pixel.  The appearance of a sequence of short (\si{\micro s} scale) \sym{RTD}s indicates that signal is present, as suggested in Fig.~\ref{fig:RTD}.  In the absence of signal, quiescent input current to the pixel is miniscule, ideally only from \ce{^{39}Ar} decay.  Cosmic muons and radioactivity in cryostat and detector plane materials also contribute, at a level presently undetermined.  If the quiescent ionization current is predominantly from \ce{^{39}Ar} decay, the time interval between \RS{}s will be on the scale of seconds.

The arrival of significant signal current leads to very short time intervals between \RS{}s, on the scale of \si{\micro s}.  In other words, input current and \RS time differences are inversely correlated: $I\propto1/\sym{RTD}$, where $I$ is the average current over an interval $\Delta T$ such that $I\cdot\Delta T = \int I(t)dt = \Delta Q$.  Here, signal currents are captured with fixed quantization $\Delta Q$ but varying time intervals.  The interval $\Delta T$ must of course be measured with a precision that facilitates accurate waveform reconstruction.  This places a requirement on time base frequency $F$. $F$, as discussed further below, is expected to be in the 50 - 100 MHz range.

To recapitulate, when a track segment arrives at a pixel, a sequence of rapid \RS{}s begins abruptly, continues for some period of time and then stops.  A sequence of short \sym{RTD}s essentially `digitizes' the current waveform: a sequence of varying time intervals, each of which represents a fixed charge `quantum'.  There is no differentiation of the signal current.  This novel form of waveform digitization appears to offer several benefits for information capture in DUNE.

If all sources of leakage current at pixel input are vanishingly small relative to that arising within the active LAr itself, then this current provides a mechanism for absolute calibration of $\Delta Q$.  Due to ASIC manufacturing variations, $\Delta Q$ will vary from one pixel to the next, likely over a significant range for all \num{e8} pixels in one module.  But since the current from \ce{^{39}Ar} decays to each pixel is known and identical, a calibration of charge sensitivity, both absolute and relative is possible.

\subsection{The proper charge quantum}

Any pixel reaching the $\Delta Q$ value triggers a \RS.  What is the charge $\Delta Q$ needed to develop an accurate representative waveform from \sym{RTD}s?  $\Delta Q$ might plausibly be chosen to correspond to $\sim1/5$ of a track crossing the pixel centrally.  For a $4\times\SI{4}{mm^2}$ pixel this is $\sim\SI{1}{mm}$.  The charge generated in a \SI{1}{mm} track segment for a Min-I particle in LAr is $\sim\num{5000}$ electrons $=\SI{5/6}{fC}$.  Taking \SI{1}{mm} as the sample size:
\[
\Delta Q \equiv \SI{5000}{electrons}\,.
\]

As \ce{^{39}Ar} decays provide a continuous average current, the CIR block may be in any state between zero and (near) $\Delta Q$ when signal current arrives.  Prior to the arrival of signal the actual threshold for information capture is thus not $\Delta Q$ but extends from $\Delta Q$ down to nearly zero.  This unusual aspect, we argue below, should be regarded not as a bug, but as a feature.

For a Min-I track segment arriving normally and assuming an electron drift velocity in LAr of \SI{1.6}{mm/\micro s}, a `waveform' sample would be taken on average each \SI{0.63}{\micro s}.  For this (atypical) track segment the instantaneous current $I$ is \SI{1.3}{nA}.  For tracks arriving parallel to the pixel plane, the instantaneous signal is spread by diffusion but will typically be $\sim5$ times larger.  In either case several waveform samples (i.e.\ \sym{RTD}s) per \si{\micro s} would occur.  For any angle of incidence, however, the profile in space-time for any track falling appreciably on a pixel will captured in good detail.  This capability could become of paramount importance if photon detection efficiency falls short of expectations, limiting event placement in drift coordinate.

\subsection{Maximum recordable current}

The shortest possible time difference between two recognizable \RS{}s establishes, at first glance, the maximum recordable current; this might be an \sym{RTD} of 2.  Is this the maximum dynamic range?  Perhaps not.  The reset circuit could be designed, at least to a good approximation, to remove precisely $\Delta Q$ per \RS.  In the event of instantaneous current much higher than $\Delta Q/t_0$, where $t_0=1/F$, a continuous sequence of the shortest possible \sym{RTD}s would occur as the input charge is nibbled down to a value that leads to an \sym{RTD} larger than the minimum.  This should allow, at least to a good approximation, a greatly increased dynamic range.  Corrections will be substantial, but are still likely of adequate accuracy.  In other words, the total integrated charge for extremely highly ionizing track segments should be rather well captured although instantaneous currents above $\sim\SI{20}{nA}$ are not as accurately represented in time. In any case, extremely highly ionizing nuclear fragments generally do not produce distinct tracks.

The \RS interval $\Delta T$ is assumed to require only a single clock tick: \SI{20}{ns}.  The instantaneous reset current needed is tiny, $\SI{1}{fC/20ns}=\SI{0.05}{\micro A}$; baseline is restored in a few \si{ns}.  If realized in practice, the magnitude of a charge loss correction for min-I particles is $\sim\SI{1}{\percent}$.

An important approach to this issue in the context of \NP has been studied by Mitch Newcomer (private communication).  In that study, an additional switch was interposed between the CSA and the pixel current input electrode.  The additional switch is normally on, allowing integration to proceed, but during \RS, this switch is momentarily turned off to prevent charge dumping during the \RS transition.  The presence of the switch may introduce fixed pattern noise/offset, but will merit further study as the best mitigation for dynamic range limitations, should that remain an issue.

\subsection{Noise, time resolution and frequency}

To provide good track profiles for true signals as well as to eliminate spurious transitions, total electronic charge noise, $\delta Q$, should not exceed about 300 electrons rms.  If achieved, the signal to noise ratio $\Delta Q/\delta Q$ per sample is approximately 17.  CIR circuitry charge noise will be manifest as fluctuations in \RS time.  Jitter in timing, $\delta t$, at the moment of \RS is given by $\delta Q/I$, where $I$ is current at input.  Thus $\delta t\propto\delta Q/I$ determines a maximum frequency $F_\text{max}\propto1/\delta t$, beyond which, for a particular instantaneous current, gain in timing precision with higher $F$ is marginal.  During the appearance of signal, these $\delta t$ fluctuations will be small, while in the absence of signal or pulses from \ce{^{39}Ar} decay, $\delta t$ will be very large.

For typical minimum-ionizing tracks with ionization signal current of \SI{1}{nA}, timing jitter $\delta t$ is $\sim\SI{50}{ns}$ rms.  For more highly ionizing tracks $F_\text{max}$ is correspondingly larger.  We argue that extremely precise $dE/dx$ measurements for the most highly ionizing tracks is unlikely to contribute usefully to physical interpretation.  Even though \sym{RTD}s for these are small, the total charge should be well measured.  We choose $F=\SI{50}{MHz}$.  This value provides very good  capture for all tracks except nuclear fragments, which in any case, do not generally produce significant tracks.

\subsection{Amplifier characteristics}

A high transconductance of at least \SI{0.1}{V/fC} is needed.  The small pixel capacitance facilitates high amplifier gain and low noise for small internal transistor currents.  The reset transistor, as indicated in Fig.~\ref{fig:CSArstSchmitt}, operates at LAr temperature and must quiescently have effectively infinite resistance to facilitate the use of current from \ce{^{39}Ar} decays for calibration.  For the same reason, the input transistor of the CSA must have negligible leakage current across its gate.  These are critical requirements for \NP.

\subsection{Nonlinear benefits?}

The CSA must be continuously sensitive, but perhaps the comparator design can display a useful non-linear power dissipation response (in the sense of class AB or B audio amplifiers).  Specifically, one may imagine that as the integrated charge nears the transition point, comparator current would increase to improve bandwidth and hence response time.  After a \RS transition, internal currents could remain high for a few \si{\micro s} to enable high-bandwidth response if the transition is the precursor to an extended series of rapid \RS (a track!).  If no track is present, the circuit would relax in a few \si{\micro s} to the ultra-low power state.

\section{Time stamping}

The most conventional approach is to distribute master clock timing signals in a cascading hierarchy down to the pixel level.  A phase-locked loop for control allows use of a lower frequency to reduce power and noise.  But nasty coherent noise from the `space-filling' time distribution network can show up only when a substantial portion of the system is realized.  While there can be little doubt that this conventional approach can be made to work, we nevertheless persist here in considering an unorthodox approach for time-stamping because it completely avoids this SPF risk, offers a flat electronic architecture and eliminates a hardware layer.

\subsection{Local Clock}

Instead of a distributed master clock, we choose to implement a local clock within the ASIC.  The local clock is based on an oscillator free-running at its own natural frequency $F$, $\pm$ some design spec, likely a few \si{\percent}.  The local oscillator is expected to run at $F=50\text{--}\SI{100}{MHz}$.  As power dissipation for this circuit element is linearly proportional to $F$, an optimization here is important.

A running calibration will be maintained, with the update rate determined by drift and dead-time considerations.  In other words, the time-stamping function is devolved maximally, to the extreme local level.  This is the approach which has been proven to be very successful in IceCube.  We assert that the problem of calibration of many local clocks is a much better problem to have than the distribution of a master clock, which presents an intrinsic and significant SPF risk.

\begin{figure}[!htb]
  \centering
  \includegraphics[width=0.9\linewidth]{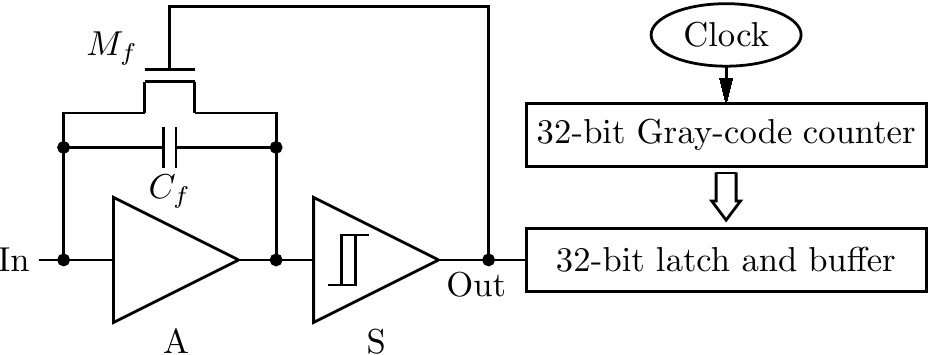}
    \caption{A free-running oscillator increments a Gray-code counter to create a local clock.  The oscillator frequency is \SI{50}{MHz}.  A \RS signal causes the local clock value to be stored in a buffer register.  Data is a string of clock snapshots, from which the \sym{RTD} is calculated by off-detector computers.}
  \label{fig:CSArstSchmittCntr}
\end{figure}

The instantaneous value of the local clock is captured in a buffer register when a \RS transition occurs among one or more of its cluster of pixels.  This is schematically illustrated in Fig.~\ref{fig:CSArstSchmittCntr}.  The string of \RS times from the free-running local clock are periodically transmitted to computers outside but near the cryostat.  The calibration task is a simple linear transformation from local clock frequencies to a central master clock.  The \sym{RTD}s are derived simply by sequential subtraction of calibrated \RS times.  At \SI{100}{MHz}, a 32-bit clock rolls over in \SI{43}{seconds}, which turns out to be more than sufficient for the DAQ function.

This scenario may seem clumsy, but works extremely well in IceCube; the process is reliable, efficient and invisible.  In IceCube, Digital Optical Modules embedded within a cubic \si{km} of ice are calibrated relatively to $\pm\SI{2}{ns}$ by this method, including electrical lengths of \si{km}-scale twisted quad cables.  This remarkably successful process occurs as a separate thread; everyone now just accepts this as `natural'.  The essential insight is that interrogation of the local clock by `surface' systems need occur only as frequently necessary to monitor and correct for oscillator drift, so that the timing requirement is met.  Here, the timing requirement, $\pm\SI{1}{\micro s}$ global accuracy corresponding to $\sim\SI{1.5}{mm}$ in the drift dimension, is 1000 times less demanding than for IceCube.

In addition, the interrogation rate must be sufficiently low that dead-time for clock calibration is negligible.  This, in turn, implies stability requirements for the local oscillators.  For example, if local clock stability is \SI{1}{ppm} over a second, then interrogation at $\sim\SI{1}{Hz}$ will suffice for $\sim\SI{1}{\micro s}$ accuracy; stability at \SI{0.1}{ppm} implies a \SI{10}{second} interval between interrogations.  If drift behavior is smooth, a fit to extended contiguous data would of course permit a lower interrogation rate.

We are confident---although we cannot yet offer evidence---that under the stable temperature conditions of LAr and nearly constant and minuscule power dissipation, a requirement of $\delta F/F < \SI{1}{ppm/sec}$ can be met.  The local clock might be a simple ring oscillator based on starved gates; other approaches are also possible.

\section{Waveforms: conversion from charge to current}

\begin{figure}[!htb]
  \centering
  \includegraphics[width=\linewidth]{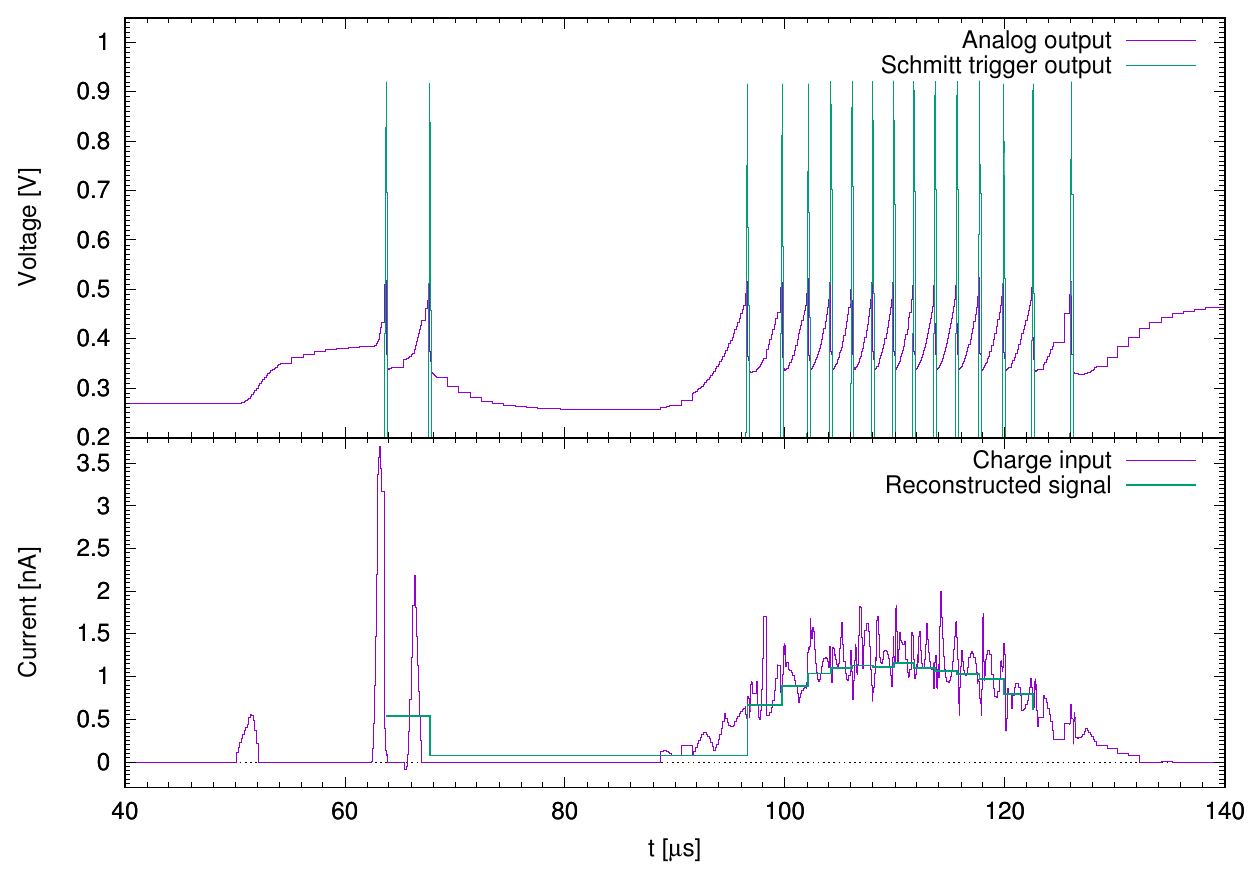}
    \caption{Transistor-level Charge Integration simulation results for a generic Min-I track current in LAr.  In this simulation of Charge-Integrate/Reset, $\Delta Q$ was chosen to be \SI{1}{fC}.  The top panel shows charge integration in purple, \RS transitions in green.  In the bottom panel, The input current (purple) and reconstructed waveform (green) is seen to be a good match.  The beginning and end of the current waveform are less well captured, but can be recovered by fitting.}
  \label{fig:CSArstFE1}
\end{figure}

\begin{figure}[!htb]
  \centering
  \includegraphics[width=\linewidth]{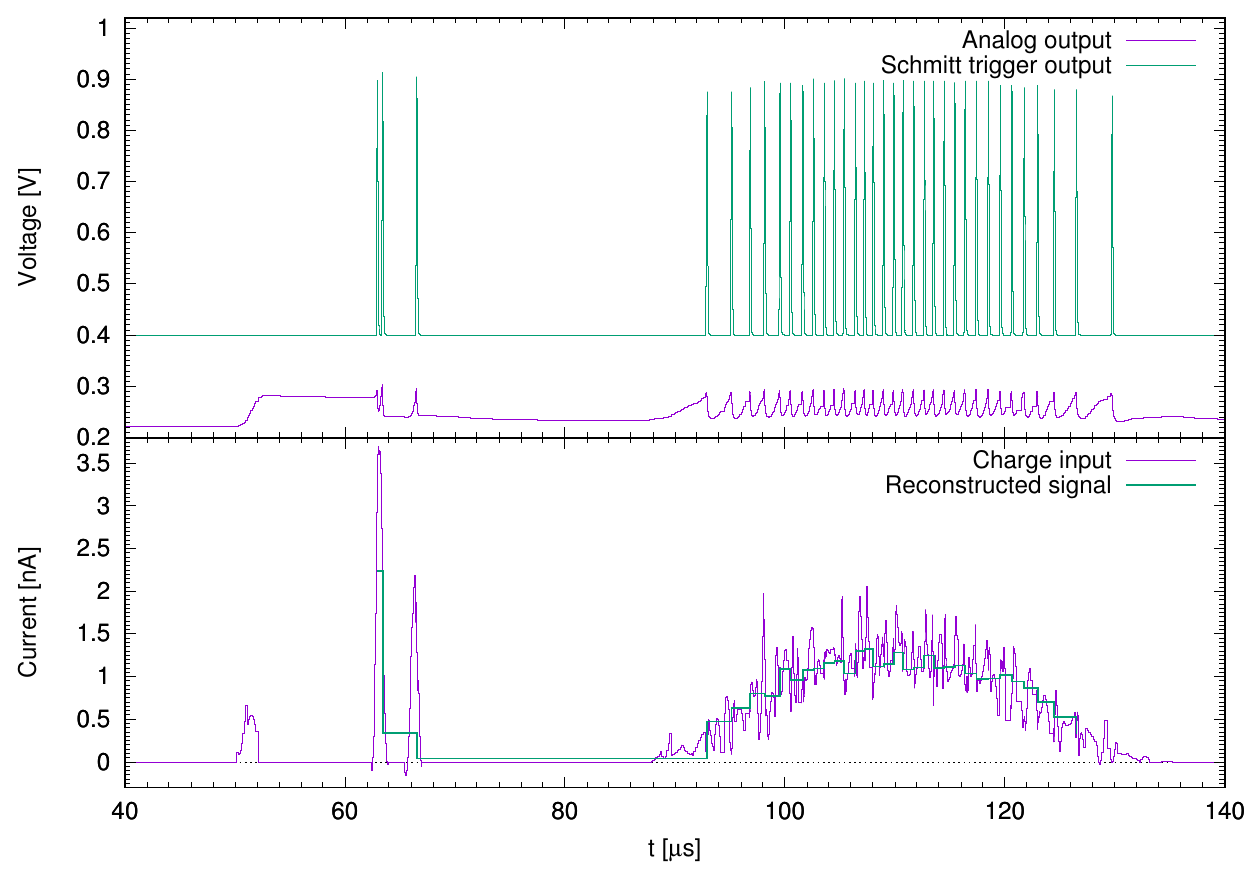}
    \caption{Transistor-level Charge Integration simulation results for a generic Min-I track current in LAr.  In this simulation of Charge Integrate-Reset, $\Delta Q$ was chosen to be $\sim\SI{0.3}{fC}$.  The top panel shows CSA integration in purple, \RS transitions in green.  In the bottom panel, The input current (purple) and reconstructed waveform (green) are a good match.  The beginning and end of the current waveform are also well captured but at the cost of very low threshold.}
  \label{fig:CSArstFE0}
\end{figure}

A contiguous sequence of short \sym{RTD}s indicates the presence of interesting signal.  A contiguous sequence of short \sym{RTD}s is the `waveform'; A conventional current waveform can be directly reconstructed.  The absence of differentiation allows current waveforms of arbitrary length and character to be captured without correlated distortion/loss/baseline droop.  This waveform, although unorthodox, can provide a complete measure of any signal of arbitrary complexity and time duration.  The input current is the same waveform in these figures, but $\Delta Q$ differs by a factor of $\sim3$.  See Fig.~\ref{fig:CSArstFE1}, \ref{fig:CSArstFE0}.  The figures suggest that an optimum value for $\Delta Q$ likely lies between 0.3 and \SI{1.0}{fC}, perhaps near \SI{0.7}{fC}.

From our idealized perspective, charge per sample is `fixed and exact' and uncertainties exist mainly in time stamping.  But even for the very largest signal currents, where the \sym{RTD} is very small---although with losses due to \RS---the absolute timing error is tiny, $\sim\SI{50}{ns}$.

The unorthodox $Q(t)\propto1/\sym{RTD}$ relationship means that larger signals will have less apparent `resolution'---but as charge per RTD is fixed, only `resolution' in time varies.  Smaller current signals will have `more', contrary to the conventional ADC.  A typical signal pulse leads to \RS after one \si{\micro s} integration and about 50 clock ticks are recorded for that waveform segment.  The track segment time is quite accurately obtained from a fit to \sym{RTD} sequence or track profile.

Due to the time-varying background current from \ce{^{39}Ar}, an uncertainty exists in the level of charge present just before the rapidly changing sequence begins.  Similarly, the very last sub-threshold sample must wait for background current to induce \RS.  But a partial correction may be inferred in either case.

For pixels adjacent to one receiving charge, there will be a small bipolar induced signal of net zero charge as charge is collected.  This well understood neighbor effect could lead to premature, false triggers for pixels nearly ready to trigger.  But, due to diffusion, this premature or false trigger seems likely to be a small effect.

\section{Auto-calibration with \ce{^{39}Ar} decay}

Aside from occasional signal, the abiding current is due to \ce{^{39}Ar}.  These $\beta$-decays lead to a steady flow of low-energy pulses plus an approximately steady current due to electron attachment.  The electron track length of any \ce{^{39}Ar} decay is less than \SI{3}{mm}.  In the charge-integrate/reset electronic architecture proposed here, with an assumption that input stage leakage current is negligible relative to \ce{^{39}Ar} decay current, the automatic detection of this current can provide an essential calibration function.  Instead of an unwanted vice, \ce{^{39}Ar} decays can be regarded as a major virtue, generating regular ``heartbeat'' \RS pulses.

The charge quantum $\Delta Q$ corresponds to $\sim1/4$ the endpoint energy loss of $\beta$ particles in \ce{^{39}Ar} decay, $\sim\SI{0.6}{MeV}$.  $\Delta Q$ is close to the average energy generated in these decays.  Each pixel views a volume of $4\times4\times7000\approx\SI{1.1e5}{mm^3}$, or about \SI{0.11}{liters}.  Assuming one \ce{^{39}Ar} decay per second per liter of LAr, one charge `pulse' occurs generally every \SI{9}{seconds} within the column viewed by one pixel.  Significant sharing among neighboring pixels from diffusion does not affect average current.

The average \ce{^{39}Ar} background current is tiny, about \SI{100}{aA}.  Since other input leakage currents must be negligible relative to this value, the \NP scheme can only work at cryogenic temperatures, and may also require special dielectric surfaces and ASIC design features and/or packaging.

\subsection{Electron attachment}

Negative ions due to attachment arrive as an approximately DC current.  For example, if average signal loss due to attachment is taken as \SI{40}{\percent}, then \SI{40}{\percent} of the total time-averaged current arriving at the pixel is in the form of negative ions, whereas the remainder arrives as signal pulses of varying amplitude down to zero.  The attachment process tends to `paint' the drift field lines with negative ions.  Many of these slowly moving `stripes' will overlap during the long ionic drift time.  Lumpiness seems likely to endure.  Recombination of negative ions with positive ions during drift is unlikely.  So the input current is a combination of approximately DC negative ion current and varying charge pulses from \ce{^{39}Ar} decay and other rare incidental processes from $\gamma$-rays, neutrinos, muons, etc.  Prior to neutralization, negative ions congregate at the metallic pixel anode surface.  Such a charge sheath will increase the local electric field.  The enhanced electric field will augment tunneling and thereby hasten neutralization (see electrostatics, below).

\subsection{Heartbeat}

The average current from \ce{^{39}Ar} decay can be expected to be quite stable over a long period and independent of attachment level.  Every live pixel thus maintains an individual but statistically regular `heartbeat' at about \SI{11}{bpm}.  Due to ASIC manufacturing variations, individual pixel feedback capacitance and the difference in \RS baseline and \RS threshold voltages will vary, and significant differences in heartbeat rate among pixels is expected.  Since offsets and capacitances are constant at LAr temperature, they will have no adverse effect on stability of the clock or heartbeat rate of a given pixel.  The heartbeats thus calibrate charge sensitivity automatically.

\subsection{\sym{RTD} distribution}

The \sym{RTD} spectrum can reveal much about free electron lifetime.  Even with electronegative impurities attaching perhaps up to \SI{40}{\percent} of the charge for maximum drift, the \sym{RTD} spectrum will be extremely bi-modal, with a histogram of broadly peaked large \sym{RTD}s (seconds \textrightarrow{} no signal) with a tail out to the largest \sym{RTD} defining an \sym{RTD}$_\text{max}$, plus a narrow peak at very much lower \sym{RTD} (\si{\micro s} \textrightarrow{} signal) values.  The histogram of \sym{RTD}s provides automatically a excellent but complex measure of pixel stability and performance.  If a pixel stops producing heartbeats, that pixel is likely dead.  The shape of the \sym{RTD} distribution should also provide an accurate dynamic measure of the liquid purity during commissioning and operation.

The \sym{RTD} distribution is also affected by the unknown amount of \ce{^{39}Ar} background charge present in the CSA prior to the appearance of true signal current.  This will lead to a sprinkling of intermediate \sym{RTD} values.  For pixels whose CSA state is near \RS threshold, charge impulses much smaller than $\Delta Q$ will trigger a transition.

A useful perspective for this is that a \SI{10}{kton} LAr detector equipped with \NP has, at any instant, approximately one kiloton of LAr with a threshold of about \SI{10}{keVee}.  Or, \SI{100}{tons} would be continually viewed with thresholds near \SI{1}{keVee}.  In this fanciful scenario, the detector could capture an intense wave of WIMPs passing through, or would be sensitive to the annual modulation of WIMP velocities.

\subsection{Hydrodynamics?}

Large-scale time-varying LAr mass-flow convection currents could be present in the detector, carrying negative ions with it and deflecting flows.  This would alter the average input DC current and affect \sym{RTD} distributions, providing a measure of time-dependent LAr hydrodynamics while possibly compromising charge calibration.

\subsection{Summary}

Overall, it is plausible to argue that the current due to \ce{^{39}Ar} decays, pulses and quasi-DC current, provides an automatic charge sensitivity calibration for each pixel.

\section{Electronic considerations and ASIC design}
\subsection{Architectural structure---physical and electronic}

\begin{figure}[!htb]
  \centering
  \includegraphics[width=\linewidth]{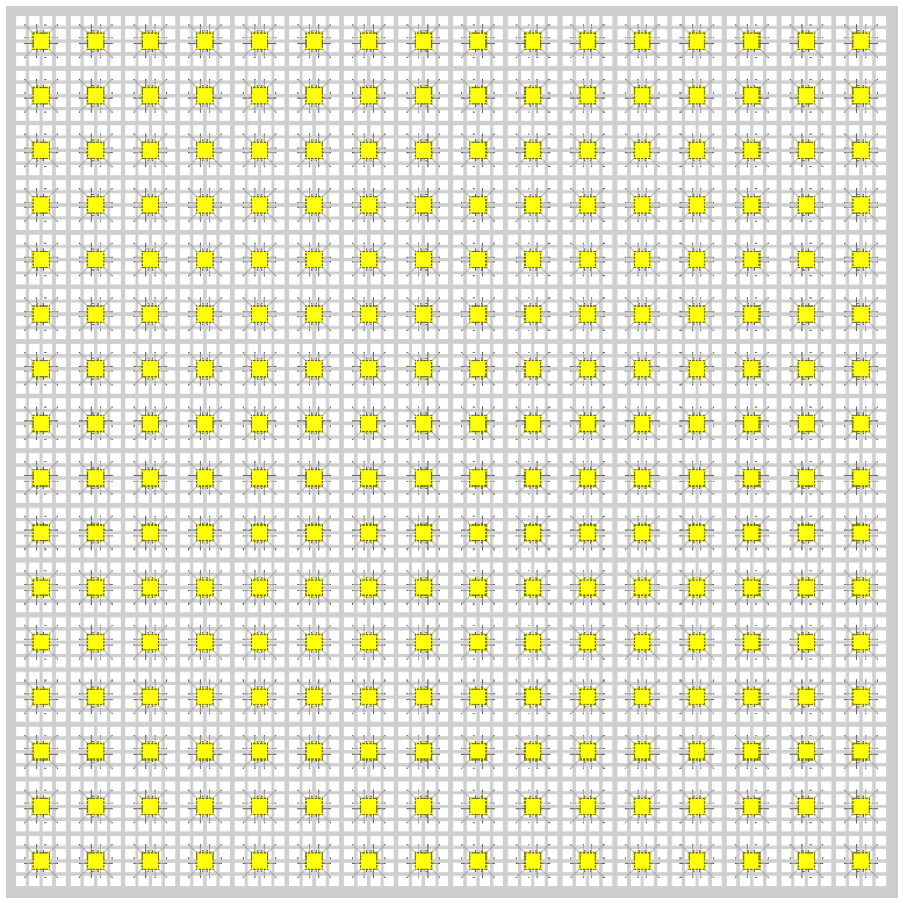}
  \caption{A $16\times16$ array a $4\times4$ pixel sub-arrays forms the `\TL' of 4092 pixels.  Local time capture and data transfer can naturally proceed from any of the four corners, or even any edge ASIC, offering a very high level of SPF protection.  This illustration is suggestive of chip-on-board but packaged chips may be better.}
  \label{fig:tile}
\end{figure}

For simplicity, and for resilience against SPF, which must be built in during design, a flat architecture is desired.  Although the ASIC is itself a genuine if elementary system-on-a-chip, a more fundamental system block is the \TL, defined as a $64\times64= 4096$ pixel block (see Fig.~\ref{fig:tile}).  With the assumption of $4\times\SI{4}{mm^2}$ pixels, the \TL is $\SI{256}{mm}\times\SI{256}{mm}$, a natural physical scale for production, handling and integration.  The organization of the \TL is conceived to be as indifferent as possible to any of various possible failure modes in any of the 256 ASICs, even in the cases where several ASICs manifest partial or catastrophic faults or display intermittencies.  The internal network of the ASICs within a \TL is dynamic and intrinsically fault-tolerant.

Only one ASIC type exists on a \TL.  Each ASIC serves 16 pixels (or perhaps 32) and comprises a complete, semi-autonomous data acquisition block.  ASIC has signal sensing, triggering, local clock, time-stamping, buffering, I/O and state machine capability.  If ASIC serves 16 pixels, 256 ASICs will be needed; for 32 pixels per ASIC just 128 ASICs will be present.

In the case of 32 pixels per ASIC, ASIC could be divided into two halves, requiring one bit to identify which half is being read out.  In either case, 8 bits are required to specify ASIC position within a \TL.  Sixteen bits are needed to flag which pixel or set of pixels within its group of 16 initiated the \RS (more than one simultaneous \RS is likely).  Tracks falling simultaneously on adjacent 16-pixel groups do not interfere.

The local clock is required to track events beyond the longest plausible time between clock/data interrogations.  At \SI{100}{MHz}, a register width of 32 bits provides 43 seconds before wrap-around.  This is more than sufficient because interrogation occurs at least once every 10 seconds.

Eight bits remain unspecified for the moment and are available as tags or other purposes.  ASIC senses all \RS transition pulses and records 32-bit time information along with pixel tags.  ASIC makes no logical decisions about meanings of a datum.  All data, whether individual heartbeats, ionization signals or timing interrogations are 64 bits and have identical structure.  All analog functions are contained in the CIR circuit block (see Fig.~\ref{fig:CSArstSchmitt}).  ASIC is connected by the shortest possible wire-bonds and traces to a $4\times4$ (or perhaps $4\times8$) sub-array of pixels.  Ionization currents arrive to a small metallic `button' exposed to the active LAr, and are transported by direct via through a dielectric layer to traces.  The button is on the scale of \SI{0.1}{mm} diameter, much smaller than the $4\times\SI{4}{mm^2}$ pixel itself.

No FPGA needs to lurk within the LAr, as data rates off-plane are remarkably low.  ASIC simply fills its buffer with clock values and tags until a clock interrogation message arrives bringing a `time-stamp token'.  At that point, the instantaneous clock value is logged with appropriate tag, and an output sequence starts.

The surface computer systems can easily recognize the heartbeat \sym{RTD}s and enter them into histograms, one for each pixel in the detector.  Timing data is extracted to that thread while re-timed \sym{RTD}s for signals are passed on to surface computers for waveform reconstruction and subsequent event building.  Only data corresponding to track ionization arrives at the surface.  The energy threshold for pulse detection varies from near zero, when charge on a pixel is high due to accumulated \ce{^{39}Ar} current, to about \SI{120}{keV}, just after \RS.  This leads to degraded sample information for the beginning and end of a typical waveform, but, these are the smallest contributions and can be reconstructed from the waveform profile with good accuracy, or possibly ignored.  Any system with a defined threshold will miss signals smaller, but here, the threshold extends down to practically zero.

ASIC design likely will employ a low-cost foundry process away from the bleeding edge for acceptable costs.  While ASIC design is complicated by the paucity of transistor/device models at LAr temperature, internal DACs in prototypes will allow needed bias optimization.  The oscillator/clock must be tightly confined within the ASIC to minimize noise propagation and power consumption.  The major engineering challenge may well be \RS transition parasitic couplings that may cause transitions in other nearby pixels.  In addition to minimized parasitic couplings, the input traces to the CIR must obviously be as short as possible to realize noise goals.

\subsection{Power dissipation}

Power dissipation and electronic noise are crucial performance desiderata.  Power dissipation must likely be less than \SI{6}{W/m^2}, hence held to less than \SI{100}{\micro W} per pixel.  The Berkeley/LBNL work shows that power dissipation, with good noise characteristics, can be in the range of few tens of \si{\micro W/pixel}.  For the local clock in \NP, which \emph{a fortiori} runs all the time, the power dissipation will approximate $(1/2)CV^2F$, where $C$ is the capacitance in silicon, $V$ is supply voltage and $F$ is the clock frequency.  This implies the total $C$ involved in the clock function cannot exceed about $\SI{1/2}{pF}$.  Our preliminary studies of power dissipation for a 32-bit Gray code clock operating at \SI{100}{MHz} in a \SI{55}{nm} CMOS process indicate a power dissipation of \SI{20}{\micro W}.  This is further encouragement that \NP may be able to meet all technical requirements.

Quiescent power dissipation occurs in three components: 1) the front-end charge integration; 2) in the quiescent current of the regenerative comparator (Schmitt trigger); 3) in the shared local free-running clock.  For the CIR block, it seems clear that power dissipation is acceptable.  For the comparator, with momentary positive feedback and high instantaneous current in the comparator, the time-averaged fraction of \RS transition is negligible.

\section{DAQ, state machine and network}

Within the \TL, all ASICs will be in one of six states, except for brief moments of state transition.  The transition period between states contributes negligibly to dead-time.  A major feature of \NP is dynamic network generation within a \TL.  This architecture faces the issue of clock domain crossing, since each ASIC clock has no relation to another; Gray code is a natural communication protocol here.  A transfer clock, if needed, must be that of the ASIC performing transfer into a neighbor ASIC's buffer.  Completion of transfer releases the transfer clock from the neighbor's buffer.  Most of the time is spent in data acquisition (DAQ).  The order here is only suggestive of normal state transitions. The allowed states are:
\begin{description}
\item[A.\ Data Acquisition (DAQ)]
Each ASIC records CIR transitions and adds 64-bit data to its buffer, without reference to activity in any other ASIC.  The duration of the DAQ state depends directly on short-term clock stability, but is expected to last normally at least one second, but likely not more than 10.

\item[B.\ Local Time Capture (LTC)]
A transition to this state begins with the introduction of an accurately timed `time stamp token' at a chosen place on the periphery of the \TL.  More than one available entry point is foreseen to reduce SPF risk.  From Fig.~\ref{fig:tile}, a clock interrogation signal---the time token---is applied to a peripheral ASIC.  This first ASIC to receive the token then asserts the token in a defined sequence to all of its other x-y neighbors; in principle, up to three neighbors could accept.

\item[C.\ Wave Propagation (WP)]
Subsequently, a possible simultaneous assertion to one ASIC by two neighbors is resolved by the accepting ASIC choosing just one, following a pre-programmed sequence.  Each ASIC in this chain remembers from whom it has accepted the token.  A signal introduced anywhere peripherally thus induces a wave propagating fully across the \TL.  Unresponsive chips will be bypassed by the encircling wave.  Propagation time across the \TL may take a fraction of one \si{\micro s}, an effect easily corrected for or ignored.  An intra-\TL network is thus dynamically established and maintained.  Should an ASIC fail at any time, though, a new dynamic network must, and will, automatically establish itself.  Although the network pattern itself is irrelevant, it can be recovered from the sequence of received data.

\item[D.\ Data Transfer (DT)]
As soon as the first ASIC learns that the proffered time-stamp token has been either accepted or rejected by all of its neighbors, read-out of all of its data captured since the previous time-stamp token will commence.  Data will include at least the one forced time capture caused by the time-stamp token, but all data captured due to CIR transitions (signal or \ce{^{39}Ar}) will also be pushed out.  The \TL begins a transition from the relatively brief LTC state to a data transfer DT state (even before the LTC state completes) as each ASIC attempts to push backward its data to the ASIC from whence it accepted the time-stamp token.  When an ASIC is empty after data transfer it must accept a data transfer token impressed from any neighbor (again following a sequence).

All data is pushed through the dynamically established network to complete the DT.  The DT phase reproduces the LTC wave in reverse but much more slowly.  Data are likely to be pushed through an average of perhaps 16 ASICs but it seems unlikely to be pushed through more than 32 ASICs.  While an inefficiency is present here, the data load is very small and infrequent.  Very substantial resilience and mechanical simplicity is obtained.  Off-plane data acquisition external to the cryostat determines when all ASICs have reported data, permitting transition back to DAQ through initiation of another DAQ phase.

\item[E.\ Initiate DAQ (IDAQ)]
Transition to this state is another wave rapidly propagating across the \TL.  In principle, no data need be captured or transferred by this wave.  IDAQ follows immediately after a \TL has been fully emptied through DAQ.

\item[F.\ Control State (CS)]
This state allows individual access to a single ASIC to permit adjustments to local oscillator frequency or internal biases, thresholds, etc.  All tokens except the CS state may be four bits long.  The CS state token is possibly 24 bits to allow for individual ASIC control and trim.
\end{description}

\subsection{Exception states}

Inevitably there will be several exception/error states that must be thought through carefully, with robust accommodation provided for at the design stage.  The simplest exception is a completely unresponsive ASIC.  In this case the wave passes around the unresponsive ASIC without difficulty.  Another possible error state is an ASIC that successfully enters and exits the LTC state but has become unable to complete data transfer and/or enable state transition back to DAQ state (or the reverse disease).  Also, in the event of a failure mode that leads to production of nonsense data, such as an oscillatory charge integrator, each ASIC must either possess built-in sensing for this that enables a kill switch, or be responsive to a message targeted to that ASIC.  It seems possible that both corrective scenarios can co-exist for this exception.

Perhaps the most lethal failure would be the short circuit, boiling LAr and/or causing a \TL to go offline.  Other nasty possibilities surely remain unmentioned here.  It will take time and effort, as usual, to ensure that all the many devils in such details have been exposed, vanquished and/or banished.

\subsection{Data Flow and Dead-time}

Altogether, the clock calibration routine, running once a second, perhaps less often, acquires \SI{16384}{bits/\TL}.  Of course, more data will be read out per ASIC when energetic signal events and/or \ce{^{39}Ar} events have occurred.  If inter-chip transfer can occur at \SI{50}{Mbits/s}, then this data load can be extracted with dead-time of $\sim\num{4e-4}$.  For a 4-bit transfer, the dead-time becomes proportionally less, $\sim\num{1e-4}$, and so forth.  The \TL{}s would experience clock calibration in a round robin sequence, so that less than one \TL in $\sim\num{e4}$ would be temporarily offline.

One \SI{10}{kton} detector (\SI{12}{m} height$\times$\SI{58}{m} length) requires \num{11136} \TL{}s per anode plane.  A \SI{7}{m} drift is imagined here, but does not affect the number of \TL{}s per plane.  For the single-phase implementation, a drift length of $\sim\SI{7}{m}$ permits two anode planes with one cathode plane per module and seems a reasonable choice.  Quiescent data rate is then $\num{22272}\times\SI{16384}{bits/s}\approx\SI{40.5}{Mbytes/s}$.  This rate seems easily manageable, and since almost all data is for the running calibration of clocks, all of that timing data is consumed locally near the LAr TPC itself and committed to \sym{RTD} histograms and other monitoring.  A tiny fraction of re-timed signal data for event candidates is transmitted onward for analysis.

This approach for data acquisition, as proposed here, incurs a small power penalty in that data are pushed through an average of maybe 16 chips.  The power cost depends slightly on how efficiently the wave swept through.  This seems likely to be an easily acceptable cost since the DT state is infrequent and the average number of bits/(second-\TL) is small.

\section{Some system considerations}
\subsection{Reliability---QA}

The charge–integrate/reset signal capture concept explicitly places essential electronic functions permanently within the LAr itself, inaccessible for the entire experiment.  It is obvious that QA practices must ensure very high long-term reliability.  IceCube, although temperatures only reached to about \SI{-50}{\degree C}, faced a similar inaccessibility problem and met those challenges successfully.  IceCube QA emphasized quality in part lists, manufacture and assembly, without recourse to either Mil-spec or NASA-style protocols.  Operational performance at the \SI{98}{\percent} level for the ensemble of \num{5000} Digital Optical Modules (DOM) has been achieved, and half of the \SI{2}{\percent} losses are attributed to cabling and connector problems during deployment.  While the temperature of LAr is much lower than in IceCube, already there exists experience with circuitry at LAr temperatures.  The most likely issue at LAr temperatures is a lack of any annealing of damage caused by hot electrons.

\subsection{Drift direction?}

The drift direction could either be vertical as in the dual-phase scenario or in the horizontal plane and perpendicular to the beam, as in the single-phase.  In either case, large two-edge-buttable staves of pixels could be assembled from smaller 4-edge-buttable boards, $25\times\SI{25}{cm^2}$ in area.  These staves could be positioned and held in vertical slots built into the cryostat internal skeleton.  Guide rails that move the staves into place after insertion seem straightforward to incorporate.  The power dissipation, however small it might be, may still induce vertical convection currents that may need counter-flow of cold LAr on the back side.

\subsection{Near Detector?}

\NP ideas and performance could be demonstrated first in a LAr `near detector' such as the `ArgonCube' concept or similar.  This is the direction pursued by the LBNL/Bern/UTA groups.  For sure, a near detector only active during beam spill could employ more conventional electronic concepts with high momentary power dissipation.  Nonetheless, a LAr TPC near detector equipped with \NP pixel planes would demonstrate continuous sensitivity with power on, serving as an essential stepping stone for implementation at the FD.

\subsection{Costs}

At this point, no useful estimate of cost, relative or absolute, is available.  While this is a critical aspect for any kiloton LAr TPC detector concept, only a careful, detailed study of costs---R\&D, engineering, QA, integration, production, QC, commissioning, etc.–--will have genuine value.

\section{Electrostatics at the pixel}

The condition we require is that (almost) all charge shall be collected on the pixel anode; therefore, almost all drift field lines reaching the pixel area must terminate on the metallic anode.  If the pixel anode, instead of a space-filling metallic surface, could be a small `button' covering perhaps only $\sim\SI{5}{\percent}$ of the pixel area, then both pixel input capacitance and inter-pixel coupling are substantially reduced.  But when the TPC HV is first turned on to establish the drift field, a large fraction of drift field lines penetrate the first dielectric layer and terminate on some ground or power plane.

One obvious solution would be a metallic lacework at the periphery of the pixel.  This structure would be negatively biased to drive electrons that arrive within the $4\times\SI{4}{mm^2}$ pixel boundary to the `grounded' pixel anode button.  The bias voltage might be in the range of \SI{-200}{V} or so.  This metallic lacework should work well enough, but is almost certainly not necessary in our view, and may act as a charge source that exacerbates charge creep.

At LAr temperature, the solution may be straightforward and automatic.  The dielectric facing the active volume should be chosen to have essentially ``infinite'' surface and volume resistivity.  In this perspective, with drift field on, the pixel dielectric surface will charge up to reach a new electrostatic equilibrium.  If field lines near the dielectric surface are not parallel to the dielectric surface, then charge drifting in will move to that insulating surface and quench that field line.  Non-parallel field lines will be almost completely eliminated at electrostatic equilibrium.  Almost all field lines from the drift region will eventually terminate on the metallic pixel button.

The \ce{^{39}Ar} current supplies about \SI{400}{fC/hour} to each pixel.  A rough estimate of the surface charge needed to approach drift field-on electrostatic equilibrium is about \SI{1000}{fC}, so equilibrium should be reached in a few hours after drift field application.  Here, ``infinite resistivity'' implies a relaxation time on the order of days; this should be possible to realize at LAr temperature with ordinary dielectric materials such as Teflon.  So the desired electrostatic result may be automatic.

Random pulses due to charge creep along the dielectric surface to pixel input at LAr temperature are a concern and would be problematic if excessive.  These events will likely be very large pulses and not resemble physics events of interest.  To avoid undesirable electric field enhancement at the anode due to surface asperities incurred during PCB manufacture, `electroless' gold plating (EGP) of the pixel button may be valuable, as EGP creates an extremely smooth surface even at atomic scale.

The possibility to use a small metallic anode button exposes most of the dielectric surface directly to LAr.  In other words, perhaps \SI{95}{\percent} of the entire anode plane is ``available'' for some other purpose: might that purpose be photon detection?

\section{Photon detection}

The detection of the extreme VUV scintillation ($\lambda\approx\SI{128}{nm}$) in LAr facilitates placement of events within the fiducial volume, unambiguously in the absence of pile-up.  The immense scale of the DUNE far detectors leads, in any scenario, to an \sym{S1} VUV signal of weak intensity spread over a very large area.  The presence of substantial Rayleigh scattering at a scale that is small relative to FD dimensions, however, dilutes optical intensity and unavoidably degrades VUV detection efficiency.  Various scenarios are in development, requiring additional scintillator/wavelength shifter/photo-detector elements.

The realization of practical high-quality signal capture for both ionization and scintillation components in the FD remains a severe technical challenge.  The ideal concept would be an integrated tracking/photo-detector system with high performance and resilience. The challenge of the very weak \sym{S1} signal intensity at the huge anode plane surface might be met, and perhaps only, by making the a large fraction of the anode plane surface photo-sensitive.

Among the field cage structures in DUNE FD geometry, the field cage surfaces are the smallest and hence least attractive candidate surface to exploit for photon detection.  The cathode plane, at very high voltage, seems best left alone.  So the \sym{S1} photons that strike the anode plane---the pixel plane dielectric ``wall''---is the best option.

Some additional mechanism must be incorporated in \NP to generate an \sym{S1} signal. Contemporary single-photon avalanche detector (SPAD) arrays, commonly known as silicon photomultipliers (SiPM), might be integrated within the \NP ASIC, depending on foundry availability.  But the use of standard opaque chip carriers would then be prohibited, and the optical path from LAr to ASIC would require wavelength shift and multiple reflections, likely a severe overall cost in efficiency.  A separate SiPM facing LAr directly seems likely to add costs, as well as complexity and fragility in manufacturing and handling.  While conceivable, such choices seem to offer marginal performance.

But there might be additional mechanisms available.  Suppose, for consideration, that the pixel dielectric surface is coated with a photoconductor such as amorphous selenium (a-Se)\footnote{Other photoconductive materials, such as the polymers used in laser printers, may be better, but a-Se is used here as a natural exemplar.}.  At LAr temperature, we must presume that quiescent elevation of carriers to the conduction band is negligible.  The steady-state electrostatic field lies somewhere between $r^{–1}$ and $r^{–2}$ near the pixel button.  When a-Se is hit by a LAr photon, however, an electron (perhaps more than one) may be elevated to the conduction band and will move in the electric field toward the button.  It is very unlikely that any single pixel will be hit by more than one \sym{S1} photon.  But there are many pixels.

If we press our imagination a bit further, let us suppose that the field near the button is sufficiently large that a single electron reaching that zone will initiate an avalanche.  Thus one converted \sym{S1} photon may lead to an avalanche charge of at least $\Delta Q$, $\sim\num{5000}$ electrons.  That pixel will fire and record the time of \sym{S1}; a ``Hit'' is generated.  The idea is that although the probability of any pixel detecting an \sym{S1} photon is small, the ensemble of $\sim\num{e8}$ pixels can produce a statistically significant set of struck pixels with similar and usefully precise timing.  A coherent \sym{S1} signal is extracted by the appearance of a statistically significant peak in the time-flow of re-timed pixel ``Hits''.

Another fanciful scenario might be that absorption of a single photon at the pixel surface could initiate charge-jumping along the surface.  Motion of sufficient charge through sufficient potential difference will lead to a pixel trigger.  A large fraction of pixels will be close to triggering due to the \ce{^{39}Ar} current.  In this notion, we invoke the phenomenon of self-organized criticality (SOC).  A system displaying SOC has static friction greater than dynamic friction and is pushed to the point of instability by constant low-level energy input.  Here, the energy is stored by charge trapped on the surface of the pixel anode.  Charge from \ce{^{39}Ar} decays continually brings charge to the anode surface, until the surface charge is sufficient to repel the field lines and cause all drifting charge to go to the metallic button.  Dielectric properties will clearly be important for any success along this line.

The photon detection scheme suggested here does not employ \sym{RTD}s, but searches for statistically significant bumps in the flow of all retimed \RS{}s, viewed in this context as ``Hits''.  The continuous scintillation background due to \ce{^{39}Ar} decay contributes a rate of Hits that will force a threshold for detection of \sym{S1} independent of any photon detection scheme.  There is likely to be a further contribution of VUV from neutralization of positive Argon ions and dimers at the cathode.  That cathode phenomenon might be suppressed by a semi-porous coating that encourages non-radiative de-excitation through complex formation.

\section{Perspective}

We have embraced the spirit of the Principle of Least Action, to develop for a large LAr TPC a self-calibrating approach for detailed information capture with very low SPF risk.  The CIR circuit, in concert with free-running local clocks, does as little as possible until something happens, yet provides readiness for detailed capture of waveforms of arbitrary complexity.  Track profiles will provide useful measures of diffusion during drift, offering a secondary means for placement in the drift dimension.  It is continuously ready for a massive data acquisition burst should a SN event occur.  The \NP detector, based on CIR, maintains a quiescent heartbeat at $\sim\SI{e10}{ns/beat}$, but automatically responds to signal with $\sim\SI{10}{ns}$ wake-up.  Due to the charge-centric architecture, information capture is efficient.  An absolute charge calibration and monitor of detector performance is naturally obtained.

IceCube (\SI{270}[\$]{M} project cost) has shown that a very large electronic investment can be successfully deployed in an inaccessible cold environment.  IceCube is also the premier exemplar of a beneficial paradigm shift away from the conventional centralized analog signal capture (AMANDA) to a decentralized digital network-based DAQ.  Perhaps something similar in impact but with very different implementation can happen here.  Even though the \NP concept for waveform capture is unfamiliar and may be regarded as a radical approach, it may become a commonplace once shown to work.  That transition happened in IceCube, as everyone now sees the DOM architecture as `obvious' even though initial resistance was quite strong.  In short, as in IceCube, if the sausage tastes good, we won't care how it is made.

The main challenge now, perhaps, is perhaps not technical, but a sociological one: to first imagine that a really good technical opportunity exists that can optimize discovery potential for DUNE, and then pursue it, strongly.

\section{Summary}

Waveform capture of the high quality described here extracts essentially all information, and may substantially extend DUNE scientific reach and discovery potential.  Two very important additional features are automatic absolute charge calibration by \ce{^{39}Ar} and robust resilience against SPF.

We wish to acknowledge particularly helpful conversations with Milind Diwan, Dan Dwyer and Mitch Newcomer.  We thank Dan Dwyer for providing simulated charge drifting signal in LAr.  We also thank Jonathan Asaadi, Jerry Blazey, Alan Bross, Eric Church, Marcel Demarteau, Zelimir Djurcic, Roxanne Guenette, Rick Van Berg, Le Xiao and Vishnu Zutshi for supportive contributions.  The authors welcome comments, criticisms, or expressions of interest.  This work was supported in part by University of Texas at Arlington Grant \#311587.  Yuan Mei is supported by a grant through Berkeley Lab, provided by the Director, Office of Science, of the U.S.\ Department of Energy under Contract No.~DE-AC02-05CH11231.



\bibliographystyle{elsarticle-num-names}
\bibliography{refs}

\end{document}